# A Missing Active Device–Trancitor for a New Paradigm of Electronics

**SUNGSIK LEE**
Department of Electronics Engineering, Pusan National University, Pusan 46241, South Korea

e-mail: sungsiklee@pusan.ac.kr

This work was supported by the Basic Science Research Program through the National Research Foundation of Korea under Grant 2018R1C1B6001688 funded by the Ministry of Science, ICT & Future Planning.

**ABSTRACT** In this paper, we first point out a missing active device while providing its theoretical definition and impact on electronics. This type of active devices has an inverse functionality of transistors and is suggested to be called trancitor rather than transistor, because it directly transfers an input signal into a voltage output. It is expected that a trancitor coupled with a transistor can provide a minimal circuit configuration, i.e., low circuit complexity, helping virtually to meet Moore's law. In addition, this may also lead to a lower power consumption and higher speed of circuits compared with a transistor-only circuit. These are supported with a circuit simulation and simple Tetris-like block analysis. In this regard, in the future, it should be required to find a trancitor to be another foundation of electronics along with transistors.

## I. INTRODUCTION

The transistor was first invented by *J. Bardeen, W. Shockley, and W. Brattain* at *Bell labs* in 1947 as the point-contact device, which was a revolutionary device to our electronics and related industries with replacing vacuum tubes [1]. And it is currently called a bipolar junction transistor (BJT) as its compact form [2]. As another type of transistors, a field-effect transistor (FET) was first patented with its concept by Linlienfeld *et al.* [3] in 1926, and was demonstrated as a metal-oxide-semiconductor FET (MOSFET) by *D. Kahng* at *Bell labs* in 1960 [4]. Since then, transistors have been evolved into various forms for a smaller scale and higher performance, including Fin-FETs and Tunneling FETs (T-FETs) [5], [6], while catching up with the Moore's law. However, the current transistor technology shows its physical limitation of down-scaling for a higher integration density, thus difficulty to satisfy with the Moore's law [7]. Since our current electronics is composed of transistors only, it is even required to employ a complicated load transistors [8]. Here, it is recognized that the load circuit is macroscopically playing the role as an inverse function of the main transistor. So, it can be argued that this usual circuit design principle with a load circuit is an alternative way in the current situation where a single active device with an inverse relationship with the transistor is missing in the elementary level.

In this paper, we introduce the missing active device, called trancitor, for the first time. This kind of active devices directly transfers an input signal into a voltage output whereas a transistor transfers its input signal into a current output, thus an inverse operation of transistors. This inverse relationship between a trancitor and transistor can naturally lead to a minimal circuit configuration as a trancitor-transistor merger in comparison with a transistor-only circuit. Eventually, it can help virtually to meet the Moore's law, guiding to a new paradigm of electronics where trancitors and transistors are combined optimally. To support these arguments, a circuit simulation and simple Tetris-like block analysis are shown.

## II. BACKGROUND
### A. EXISTING ACTIVE DEVICES - TRANSISTORS

Transistors, such as BJTs and FETs, are commonly active devices where the input signal is transferred into the current at the output, thus it works like a variable resistor [9]. So, it is called transistor as a combined terminology from transfer and resistor [10]. From a circuit theory point of view [8], the transistor is included in the category of a current source (CS) controlled by a current (I) or voltage (V) input. As seen in Fig.1, a BJT is a current-controlled current source (CCCS) with its governing equation represented as,

$$I_{out} = f_1(I_{in}) = \beta I_{in}, \tag{1}$$

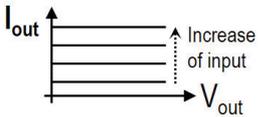

| Input Output | Current (Current-Controlled: CC) | Voltage (Voltage-Controlled: VC) |
|---|---|---|
| **Current** (Current Source: CS) 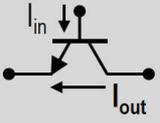 | **Bipolar Junction Transistor (BJT)** (CCCS) 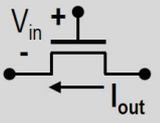 Transistor effect | **Field Effect Transistor (FET)** (VCCS) 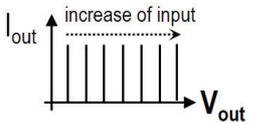 Surface charge effect |
| | **Transistor** (Transfer + Resistor) | |
| **Voltage** (Voltage Source: VS) 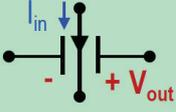 | **Missing** (CCVS) 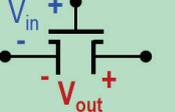 Electromagnetic effect ? (e.g. Hall-effect) | **Missing** (VCVS) Electromagnetic effect ? |
| | **Trancitor** (Transfer + Capacitor) | |

**FIGURE 1.** A theoretical list of elementary active devices deduced from 4 possible combinations of the current and voltage at the input and output, respectively. Here, it is found that trancitors are missing as CCVS or VCVS-type active devices (Inset: symbolic representations for each category).

where the output current ($I_{out}$) at the collector is a function of the input current ($I_{in}$) at the base, and $\beta$ is an amplification factor [11]. Note that a BJT can also be modeled as a charge-controlled CS or voltage-controlled CS [8]. In any case, it is still a CS device. Similarly, FETs are in the category of a voltage-controlled current source (VCCS), and the basic current formula for the saturation regime in a MOSFET, the most popular type of FETs [10], is as follows,

$$I_{out} = f_2(V_{in}) = \frac{1}{2}\frac{W}{L}K(V_{in} - V_T)^2. \quad (2)$$

Here, $I_{out}$ is a drain current as the output controlled with a gate voltage as the input ($V_{in}$), K is a constant determined with the device's physical properties of layers in a MOSFET, W/L is the ratio of channel width (W) and length (L), and $V_T$ is a threshold voltage [11]. As reflected in Eqs.1 and 2, BJTs and FETs are commonly current source (CS) elementary devices.

### B. UNRECOGNIZED ACTIVE DEVICE - TRANCITORS

Revisiting Fig.1, it is also recognized that two other active devices can naturally be deduced. However, to the best of our knowledge, current-controlled voltage source (CCVS) and voltage-controlled voltage source (VCVS) types are missing in the elementary level, which are commonly voltage source (VS) devices. Due to the absence of these devices, a transistor-based circuit alternatively employs a complicated load circuit macroscopically operating as those elementary devices [8]. And this gives rise to a high complexity and power-consumption. In other words, if there exists, the VS elementary device can directly provide a voltage output without a sophisticated load circuit, while minimizing a circuit complexity and power-consumption. This category of devices is transferring its input into a voltage signal at the output working like a capacitor. So, it can be named trancitor as a compound word of transfer and capacitor. The analytical forms of governing equations for CCVS and VCVS-type trancitors can be expressed assuming their transfer-linearity, respectively,

$$V_{out} = g_1(I_{in}) = \alpha_i I_{in}, \quad (3)$$
$$V_{out} = g_2(V_{in}) = \alpha_v V_{in}, \quad (4)$$

where $\alpha_i$ and $\alpha_v$ are transfer coefficients between the input and output in CCVS and VCVS-type trancitors, respectively. And these coefficients are determined and designed when a respective device concept and physical mechanisms are found. Note that the dimension of $\alpha_i$ is a resistance given from $V_{out} / I_{in}$ in Eq.3 while $\alpha_v$ in Eq.4 is dimensionless.

## III. THEORETICAL PERSPECTIVES

### A. POSSIBLE STRUCTURE AND OPERATING PRINCIPLE

The operating principle of a trancitor are currently unknown. However, focusing on its output voltage signal generated with its input signal, it may be governed by the Hall-effect, especially for the CCVS-type trancitor, which can induce the voltage signal called the Hall voltage [11]. As seen in Fig.2(a), the magnetic flux density ($\mathbf{B}_z$) is applied along the z-direction perpendicularly to the semiconductor film on the x-y plane.

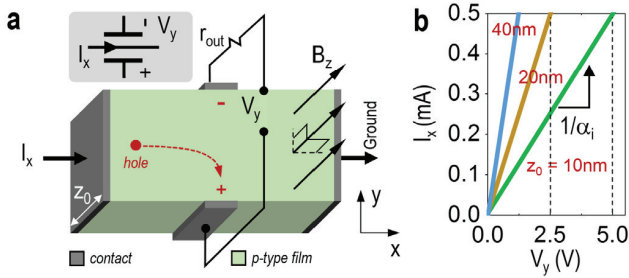

**FIGURE 2.** (a) Possible structures of trancitors, especially for the CCVS-type. (b) Simulation results of the input-output transfer characteristics for a different semiconductor thickness for $B_z = 1$ $\mu$Wb/cm$^2$ and p = $6.2 \times 10^{14}$ cm$^{-3}$.

Upon applying this magnetic field, the moving charge (Q), composing the input current (i.e. $I_x$) to the x-axis, can get the Lorentz force to the y-direction, ($\mathbf{F}_y$) as,

$$\mathbf{F}_y = Q\mathbf{E}_y = Q\mathbf{v}_x \times \mathbf{B}_z, \tag{5}$$

where $\mathbf{E}_y$ is the induced electric field intensity along the y-axis, and $\mathbf{v}_x$ is the drift velocity of the moving carriers. In Eq.5, $\mathbf{v}_x$ is defined as $x_0/t_0$, where $x_0$ is the moving distance on the x-axis and $t_0$ is the transit time. With this definition, since $I_x = Q/t_0$, $\mathbf{v}_x$ can be rewritten as follows,

$$\mathbf{v}_x \equiv I_x \frac{x_0}{Q} \hat{\mathbf{x}}. \tag{6}$$

Here, $\hat{\mathbf{x}}$ is the unit vector for the x-axis. With Eqs.(5) and (6), $\mathbf{E}_y$ is found as,

$$\mathbf{E}_y = -I_x \frac{x_0}{Q} B_z \hat{\mathbf{y}}, \tag{7}$$

where $B_z$ is the magnitude of the magnetic flux density. Now, the line integral of Eq.(7) along the y-axis gives the Hall voltage ($V_y$) as,

$$V_y = -\int_0^{y_0} \mathbf{E}_y \circ dy \; \hat{\mathbf{y}} = \frac{I_x B_z x_0 y_0}{Q}. \tag{8}$$

Here, $y_0$ is the width of the semiconductor film, $\hat{\mathbf{y}}$ is the unit vector for the y-axis. In Eq.8, if the semiconductor is a p-type, Q can be defined with the hole concentration (p) and geometry,

$$Q = q \int_v p dv \approx qpx_0 y_0 z_0, \tag{9}$$

where $z_0$ is the thickness of the film. From Eqs.8 and 9, $V_y$ is analytically given as,

$$V_y = \frac{I_x B_z}{qpz_0}. \tag{10}$$

In Eq.10, $V_y$ is $V_{out}$ while $I_x$ is $I_{in}$, thus the expression of $\alpha_i$ is also found as,

$$\frac{V_y}{I_x} = \frac{V_{out}}{I_{in}} \equiv \alpha_i = \frac{B_z}{qpz_0}. \tag{11}$$

### B. TRANCITOR-TRANSISTOR COMBINATION

Among those missing devices, particularly, the CCVS-type trancitor can be paired with the MOSFET (VCCS) since they are in an inverse relationship with each other. This implies that they can provide a minimal configuration of their circuit. In other words, the trancitor-transistor merger can lead to a lower circuit complexity compared to a transistor-only configuration. For example, a voltage amplifier can be designed with one trancitor and transistor. And its characteristic equation is easily given from the compositional function of Eqs.2 and 3, as,

$$V_{out} = g_1\{f_2(V_{in})\} = \frac{1}{2} \frac{W}{L} \alpha_i K (V_{in} - V_T)^2. \tag{12}$$

In addition, the voltage gain ($A_v$) is derived from the first derivative of Eq.12 with respect to $V_{in}$, as follows,

$$A_v = \frac{W}{L} \alpha_i K (V_{in} - V_T). \tag{13}$$

To support these arguments, a circuit simulation is performed employing a standard MOSFET model with $K = 2 \times 10^{-5}$ C/V$^2$-s, and the CCVS-type trancitor model based on the Hall device model with $\alpha_i = 10^4 \Omega$ [12]. As shown in Figs.3a and b, while a FET-only circuit employs a complicated load circuit to convert a current signal into a voltage signal, a trancitor-transistor pair shows a minimal complexity with providing the same function as the FET-only voltage amplifier (see Fig.3c). Indeed, both circuits exhibit the voltage gain ($A_v$) of 10 at a low frequency. For $K = 2 \times 10^{-5}$ C/V$^2$-s, $W/L = 50$, and $\alpha_i = 10^4 \Omega$, the calculation with Eq.6 gives the same value of $A_v$. Note that, compared to one of the simplest versions of a homogeneous FET amplifier circuit (see Fig.3a), an operational amplifier (i.e. op-amp) may be a better example of a high complexity. Here, much more transistors are employed to achieve high gain and low output impedance [8]. Even in a common-source amplifier configuration, a current mirror circuit is actually required for biasing the active load, as seen in Figs.4(a) and (b). So, the version in Fig.3(a) is only a very simplified version. Besides, a power consumption of a trancitor-transistor circuit is also reduced due to a supply voltage ($V_{Supply}$) lowered by the absence of load transistors (see Fig.3d). As another merit, it is also found that the cut-off frequency of paring a transistor with a trancitor is higher than that of the other case based on transistors only, implying a higher operating speed (see Fig.3e). This can be explained with a lower input impedance of the trancitor load compared to the FET-only load. Here, it is expected that such a simple circuit seen in Fig.3c can play the role as a very complicated op-amp composed of much more transistors even in comparison with the circuit shown in Fig.3a.

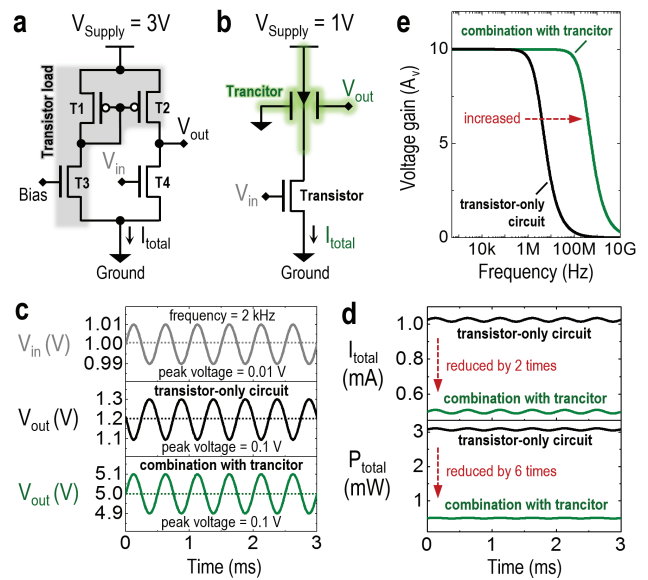

**FIGURE 3.** (a) Schematics of a simplified version of voltage amplifier only based on the least number of transistors (i.e. transistor-only circuit). Here, T1 and T2 are P-MOSFETs while T3 and T4 are N-MOSFETs. And transistors T1 to T3 compose a load circuit to have macroscopically an inverse function of the main transistor T4. (b) Circuit diagram of a trancitor-transistor combination to be a voltage amplifier with an equivalent functionality to the transistor-only circuit. (c) Transient waveforms of the input and output signals for each case. (d) Total current and power-consumption as a function of time for each circuit. (e) Plot of the voltage gain versus frequency comparing the transistor-only case with the other case where a trancitor and transistor are combined.

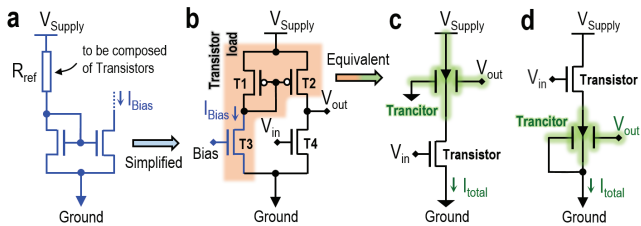

**FIGURE 4.** (a) current mirror circuit with a resistor load. Here, it is recognized that the reference resistor ($R_{ref}$) also needs to be composed of transistors as well, leading to a higher complexity. (b) A simple common source amplifier with transistors only. Here, a single transistor is simplified to be a current source. (c) Circuit with the trancitor as an equivalence to the complicated transistor load. (d) Another version of the combinational circuit of the trancitor and transistor along with the same functionality as the circuit in (b) and (c).

### C. IMPACT ON MOORE's LAW

In order to further examine the complexity of the trancitor-transistor circuit, we employ a simple Tetris-like block analysis for a possible generalization and visualization [13], while comparing it with the case of the transistor-only circuit. Here, as shown in Fig.5a, the transistor and trancitor are conceptually modelled as T and U-shaped blocks, respectively, since they are an inverse form with each other. Here, we assume that the electrical function of each device is corresponding to the geometrical form of it. Note that a validity of the provided Tetris analysis is subject to changing or improvement of a market demand and technology node. Based on the T-shaped blocks only, a square area needs four of them to fully cover itself (see Fig.5b). In contrast, merging the T-shaped block with the U-shaped block easily makes an even smaller square with just two of themselves (see Fig.5c). In this case, assuming that the critical dimension ($\lambda$) and the functionality of each device are maintained, the area consumption is reduced by more than 43% compared to the T-shaped blocks only, as indicated in Figs.5b and c. For the same area of the square, four of the T-shaped blocks need to be scaled down to fit there. Indeed, its critical dimension is shortened by 25%, as labeled in Figs.5c and d. This implies that the mixture of the U-shaped (trancitor) and T-shaped blocks (transistor) without down-scaling can make the same effect as four of transistors scaled down by 25% (i.e., reduced dimension, $\Delta = 0.25\lambda$). Here, we assume that the trancitor has the most minimal U-shape shown in Fig.5a. Note that it can be larger depending on actual technology and layout of the trancitor. With this, the increase rate in the effective density of transistors ($\zeta_{EDOT}$), is calculated with,

$$\zeta_{EDOT} = \frac{1}{(1 - \Delta/\lambda)^2}. \quad (14)$$

By Eq.14 for $\Delta = 0.25\lambda$, $\zeta_{EDOT}$ is found to be 1.78 as marked in Fig.5g. Practically, those effects would be more than those because a practical layout of transistors needs a more spacing and margins between many different layers, as seen in Fig.5e [14]. Moreover, a high performance voltage amplifier based on transistors only needs many transistors more than 4. This means that the Tetris block analysis shown here provides a worse case than the practical situation. Consequently, these results support that the incorporation of a trancitor into an existing transistor technology would help virtually to meet the Moore's law (see Fig.5g).

### D. SUITABILITY OF TRANCITORS FOR EMERGING TECHNOLOGIES

In this respect, a combination of trancitors and transistors can give a more minimal design of circuits compared to a transistor-only case for the same functionality. And its simplicity would be useful especially for thin film technologies fabricated with a low cost process of a large critical dimension [15], [16]. As discussed in the previous section, the usage of trancitors gives an effect of a 25% reduction in the feature size. At the same time, a thin-technology has natural advantages to get a larger $\alpha_i$ for a higher voltage gain in Eq.13, since it is based on a film deposition with a tiny thickness in the range of a few tens of Nano-meters. From Eq.11, its expression is recalled,

$$\alpha_i = \frac{B_z}{qpz_0}. \quad (15)$$

As can be seen in Eq.15 and Fig.2(b), the $\alpha_i$ can be larger with a smaller $z_0$, i.e. thinner semiconductor film, thus a suitability of thin film technologies for trancitors. As a numerical example, with Eq.15, $\alpha_i$ is calculated as $10^4 \Omega$ for the given parameters, such as $B_z = 1$ $\mu$Wb/cm$^2 = 0.01$ T, $p = 6.2 \times 10^{14}$ cm$^{-3}$, and semiconductor thickness ($z_0$) = 10 nm.

As seen in Fig.3d, a power consumption of the trancitor-transistor circuit has about 0.5 mW which is 6 time lower compared to the transistor-only circuit. This low power consumption of a trancitor-transistor circuit can also be an essence particularly for wearable electronic systems in the internet-of-things (IoT) where a battery lifetime should be extended [15], [17]. In addition, an operating speed of the trancitor-transistor circuit has almost two orders of magnitude faster than the transistor-only circuit, as found in Fig.3e. So, it is also expected that a trancitor-transistor pair would be suitable for a neuromorphic circuit where a low complexity and high processing speed are essential [18]. These suggest that trancitors would bring a new paradigm of electronics where trancitors and transistors are making an optimum state with their inverse relationship.

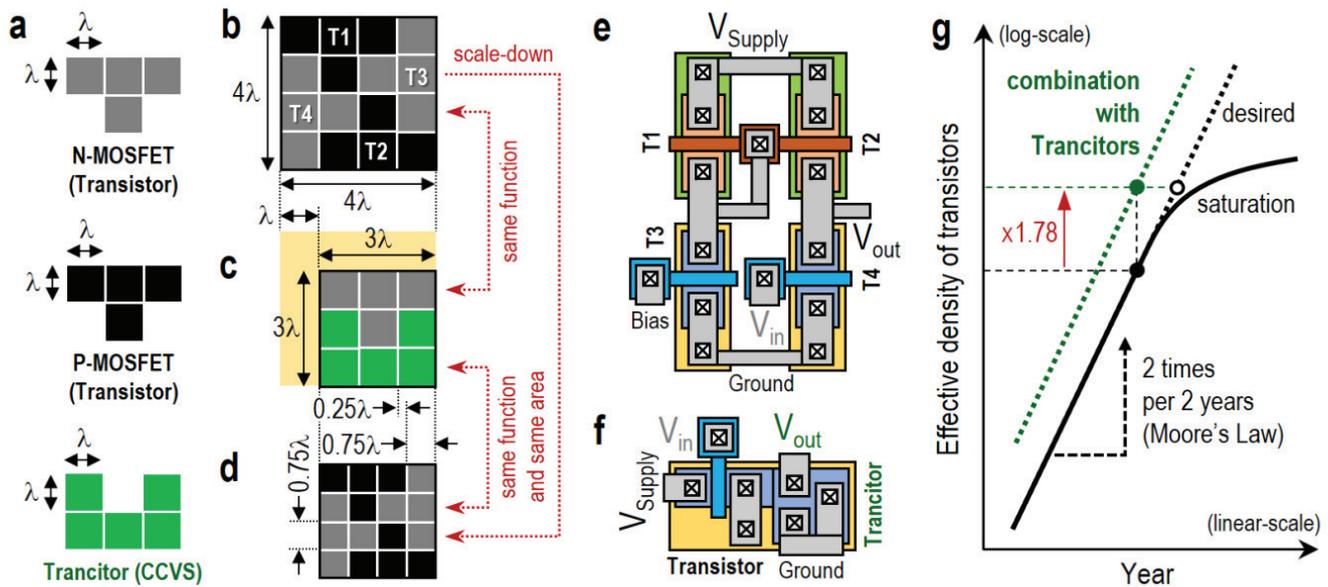

**FIGURE 5.** (a) Definition of each block. Here, the T-shaped and U-shaped blocks are representing MOSFETs and trancitors, respectively. (b) 4λ × 4λ square filled with four of T-shaped blocks. (c) 3λ × 3λ square covered with one T-shaped and U-shaped block. Here, the yellow area indicates a 43% reduction in comparison with the 4λ × 4λ square. (d) Square filled with four of T-shaped blocks scaled to be the same area as the 3λ × 3λ square. Here, the critical dimension becomes smaller by 25% compared to other cases. (e) Typical layout of the transistor-only circuit seen in Fig.4b, which is the practical case for the 4λ × 4λ square with 4 T-shaped blocks. (f) Example layout for the trancitor-transistor pair of the circuit shown in Fig.4d, representing the case of the 3λ × 3λ case. (g) Plot of the effective density of transistors versus year, where an effect of the combination with trancitors on Moore's law is indicated.

## IV. CONCLUSION
It is important to find that there is a missing elementary active-device to be called trancitor as an inverse form of a transistor. While our current state of electronics is composed of transistors only, it turns out that the absence of a trancitor makes electronic circuits very complicated with employing sophisticated loads. So, it is important to find the trancitor for a new paradigm of electronics coupled with transistors, leading to a lower complexity, lower power-consumption, and even higher operating speed of their circuit compared to a transistor-only circuit. Finally, it is strongly implied that the proposed trancitor concept may be one of the only ways to go beyond CMOS electronics, where a totally different operating principle should be applied while revolutionizing our electronics in a more fundamental way, along with trancitors, rather than just scaling down the existing transistors for meeting the Moore's law.

## ACKNOWLEDGMENT
The author would like to thank Dr. X. Cheng at Cambridge University for the useful and meaningful discussions, and tribute to Dr. D. Kahng for his earlier works on electronic devices as a motivation for this work.

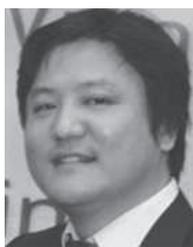

**SUNGSIK LEE** received the Ph.D. degree from University College London, London, U.K., in 2013. From 2013 to 2017, he was a Research Associate with the University of Cambridge, Cambridge, U.K.. He has been a Professor with the Department of Electronics, Pusan National University, Pusan, South Korea, since 2017. He has published over 70 articles in the field, including one journal *Science* on the almost-off transistor as the first author. His area of expertise is semiconductor devices and physics for futuristic electronics. He was a recipient of the IEEE EDS Ph.D. Student Fellowship in 2011. He received the Best Teaching Prize from the Korean Society for Engineering Education, South Korea, in 2017.

• • •